# MODELLING OF PYROELECTRIC RESPONSE IN INHOMOGENEOUS FERROELECTRIC-SEMICONDUCTOR FILMS

A. N. Morozovska[1], E.A. Eliseev[2], D. Remiens[3], C. Soyer[3]

[1]V.Lashkaryov Institute of Semiconductor Physics, NAS of Ukraine,

pr. Nauki 41, 03028 Kiev, Ukraine, e-mail: morozo@i.com.ua

[2]Institute for Problems of Materials Science, National Academy of Science of Ukraine,

3, Krjijanovskogo, 03142 Kiev, Ukraine, e-mail: eliseev@i.com.ua

[3]IEMN, UMR 8520 OAE-dept/ MIMM, Universite de Valenciennes et du Hainaut-Cambresis, Le Mont Houy, 59313 Valenciennes Cedex 9, France

## ABSTRACT

We have modified Landau-Khalatnikov approach and shown that the pyroelectric response of inhomogeneous ferroelectric-semiconductor films can be described by using six coupled equations for six order parameters: average displacement, its mean-square fluctuation and correlation with charge defects density fluctuations, average pyroelectric coefficient, its fluctuation and correlation with charge defects density fluctuations.

Coupled equations demonstrate the inhomogeneous reversal of pyroelectric response in contrast to the equations of Landau-Khalatnikov type, which describe the homogeneous reversal with the sharp pyroelectric coefficient peak near the thermodynamic coercive field value. Within the framework of our model pyroelectric hysteresis loop becomes much smoother, thinner and lower as well as pyroelectric coefficient peaks near the coercive field completely disappear under the increase of disordering caused by defects. This effect is similar to the well-known "square to slim transition" of the ferroelectric hysteresis loops in relaxor ferroelectrics. Also the increase of defect concentration leads to the drastic decrease of the coercive field typical for disordered ferroelectrics.

Usually pyroelectric hysteresis loops of doped and inhomogeneous ferroelectrics have typical smooth shape without any pyroelectric coefficient peaks and coercive field values much lower than the thermodynamic one. Therefore our approach qualitatively explains available experimental results. Rather well quantitative agreement between our modelling and typical $Pb(Zr,Ti)O_3$-film pyroelectric and ferroelectric loops has been obtained.

**Keywords:** ferroelectrics-semiconductors, charged defects, coercive field, pyroelectric hysteresis loops, coupled equations.

PACS 77.80.-e, 77.80.Dj, 61.43.-j

## 1. INTRODUCTION

The main peculiarity of ferroelectric materials is hysteresis of their dielectric permittivity $\varepsilon$, spontaneous displacement $D$ and pyroelectric coefficient $\gamma$ over electric field $E_0(t)$ applied to the sample [1]. Dielectric $\varepsilon(E_0)$, ferroelectric $D(E_0)$ and pyroelectric $\gamma(E_0)$ hysteresis loops in an inhomogeneous ferroelectric-semiconductor film have several characteristic features depicted in Fig.1a.

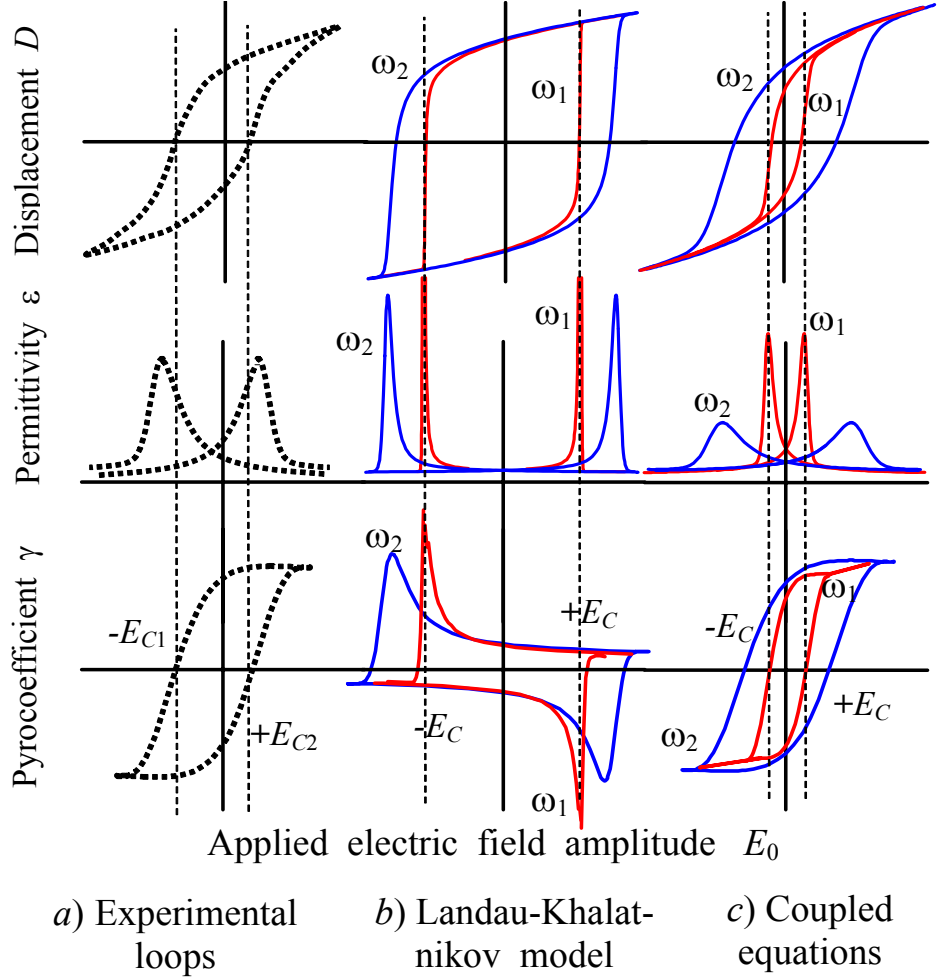

**Figure 1**. Dielectric $\varepsilon(E_0)$, ferroelectric $D(E_0)$ and pyroelectric $\gamma(E_0)$ hysteresis loops. Different plots correspond to the data obtained for a semiconductor-ferroelectric film (a), Landau-Khalatnikov model (b) and our coupled equations (c) for a bulk sample ($\omega_1 \ll \omega_2$ are two frequencies of applied electric field). Note, that Landau-Khalatnikov thermodynamic coercive field $E_C$ is at least several times greater than the experimentally observed values $E_{C1}$ and $E_{C2}$ (sometimes the loops observed in the films with thickness 50nm-5μm are shifted on the value $E_i = E_{C2} - E_{C1}$, also the shift can be time-dependent).

Nowadays such typical ferroelectric-semiconductors as $BaTiO_3$ ceramics, $Pb(Zr,Ti)O_3$ solid solutions slightly doped with Fe, Cr or La, Nb, Nd, Ce *etc.*, their films, multilayers and heterostructures are widely used in actuators, electro-optic, piezoelectric, pyroelectric sensors and memory elements [2]-[4]. However, the task how to create ferroelectric material with pre-determined dielectric and/or pyroelectric properties is solved mainly empirically. The correct theoretical consideration and modelling of related problems seems rather useful both for science and applications. It could answer fundamental questions about the nature of lattice-defects correlations, possible self-organization in the system and help to tailor new ferroelectric-semiconductor materials, save time and expenses.

Conventional phenomenological approaches [5], [6] with material parameters obtained from first-principle calculations [7], [8] can be successfully applied to the pure ferroelectrics materials [9], [10], but they give significantly incomplete picture of the dielectric, ferroelectric and pyroelectric hysteresis in the doped or inhomogeneous ferroelectrics-semiconductors (compare Fig.1b with Fig.1a).

In particular, pioneer Landau-Khalatnikov approach, evolved for the single domain pure ferroelectrics-dielectrics [11], [12], describes homogeneous polarization reversal but presents neither domain pinning nor domain nucleation and domain movement. The calculated values of thermodynamic coercive field [1] are significantly larger than its experimental values for real ferroelectrics [3], [13]. Observed hysteresis loops usually look much thinner and sloped than Landau-Khalatnikov ones [3], [14]-[17]. Also equations of Landau-Khalatnikov type describe homogeneous pyroelectric response reversal with the sharp peak near the thermodynamic coercive field value. We could not find any experiment, in which the pyroelectric coefficient peak near the coercive field had been observed. Usually the pyroelectric hysteresis loops of doped ferroelectrics have typical "ferroelectric" shape with coercive field values much lower than the thermodynamic ones. Therefore, the modification of Landau-Khalatnikov approach for inhomogeneous ferroelectrics-semiconductors seems necessary [18].

In our recent papers [19]-[23] we have considered the displacement fluctuations and extrinsic conductivity caused by charged defects and thus have modified the Landau-Khalatnikov approach for the inhomogeneous ferroelectrics-semiconductors. The derived system of three coupled equations gives the correct description of dielectric and ferroelectric hysteresis loops shape and experimentally observed coercive field values (see Fig.1c). In this paper we develop the proposed model [20] for pyroelectric response and demonstrate that the pyroelectric hysteresis loops of ferroelectrics-semiconductor films with charged defects can be successfully described by using six coupled equations.

## 2. COUPLED EQUATIONS

Let us consider $n$-type ferroelectric-semiconductor with sluggish randomly distributed defects. The charge density of defects $\rho_s(\mathbf{r})$ is characterized by the positive average value $\overline{\rho_S}$ and random spatial fluctuations $\delta\rho_S(\mathbf{r})$, i.e. $\rho_S(\mathbf{r}) = \overline{\rho_S} + \delta\rho_S(\mathbf{r})$. Hereinafter the dash designates the averaging over the sample volume. The charged defects distribution is quasi-homogeneous. The average distance between defects is $d$. Movable screening clouds $\delta n(\mathbf{r},t)$ with Debye screening radius $R_D$ surround each charged center, so free carriers charge density $n(\mathbf{r},t) = \overline{n} + \delta n(\mathbf{r},t)$ is characterized by the negative average value $\overline{n}$ and modulation $\delta n(\mathbf{r},t)$. Screening clouds are deformed in the external field $E_0$, and the system "defect center $\delta\rho_S$ + screening cloud $\delta n$" causes displacement fluctuations $\delta D(\mathbf{r},t)$ in accordance with Maxwell's equations $div\,\mathbf{D} = 4\pi(n + \rho_s)$, $div(\partial\mathbf{D}/\partial t + 4\pi\,\mathbf{j}_c) = 0$ (see Fig.2, Fig. 1 in [20] and Fig.3.2 in [5]).

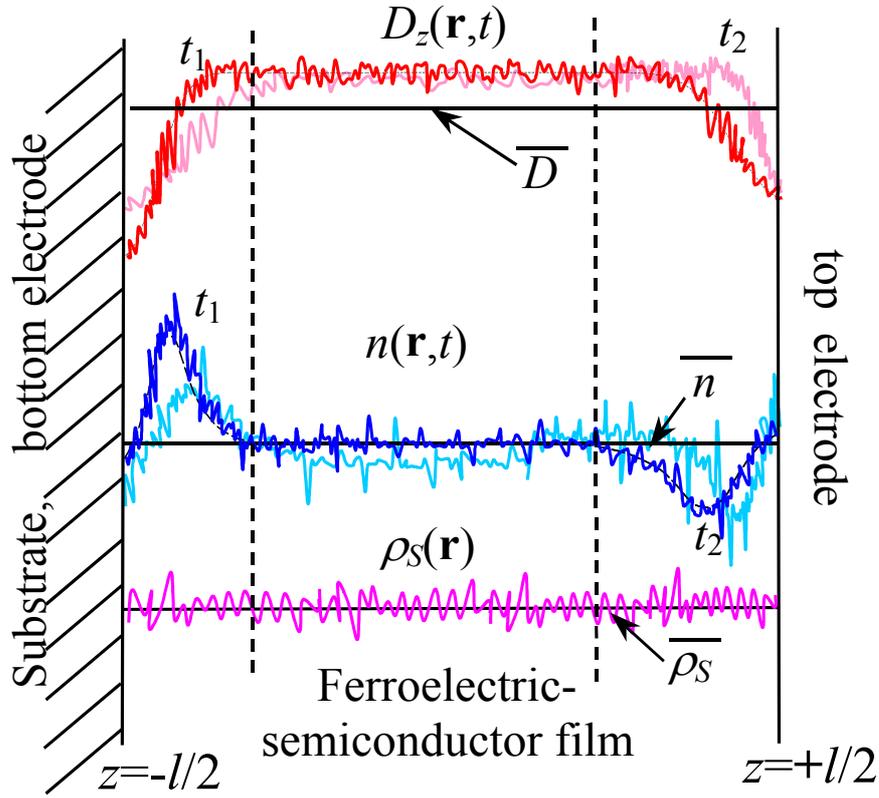

**Figure 2**. Spatial distribution of displacement $D$, free carriers charge density $n$ and sluggish defects density $\rho_S$ in an inhomogeneous ferroelectric-semiconductor film.

In our papers [20]-[22], we have modified classical Landau-Khalatnikov equation $\Gamma\,\partial D/\partial t + \alpha D + \beta D^3 - \gamma\,\partial^2 D/\partial\mathbf{r}^2 = E_z$ for bulk samples and obtain the system of coupled equations for average displacement $\overline{D}$, its mean-square fluctuation $\overline{\delta D^2}$ and correlation with charge defects density fluctuations $\overline{\delta D \, \delta \rho_S}$. The average pyroelectric coefficient $\overline{\gamma}$ can be easily obtained by the differentiation of the coupled system [20] on the temperature $T$. We suppose that only coefficient $\alpha = -\alpha_T (T_C - T)$ and screening radius $R_D \approx \sqrt{k_B T / 4\pi \overline{n} e}$ essentially depend on temperature in usual form $\partial \alpha / \partial T = \alpha_T$ [1], $\partial R_D^2 / \partial T = k_B / 4\pi \overline{n} e$ [24], [25]. In this way we obtained six coupled equations for the average displacement $\overline{D}$, its mean-square fluctuation $\overline{\delta D^2}$ and correlation $\overline{\delta D \, \delta \rho_S}$, pyroelectric coefficient $\overline{\gamma} = \partial \overline{D} / \partial T$, its deviation $\overline{\delta \gamma} = \partial \overline{\delta D^2} / \partial T$ and correlation with charge defects density fluctuations $\overline{\delta \gamma \, \delta \rho_S} = \partial \overline{\delta D \, \delta \rho_S} / \partial T$ (see Appendix A):

$$\Gamma \frac{\partial \overline{D}}{\partial t} + \left( \alpha + 3\beta \, \overline{\delta D^2} \right) \overline{D} + \beta \overline{D}^3 = E_0(t) + E_i(l,t), \tag{1}$$

$$\frac{\Gamma_R}{2} \frac{\partial \overline{\delta D^2}}{\partial t} + \left( \alpha_R + 3\beta \overline{D}^2 \right) \overline{\delta D^2} + \beta \left( \overline{\delta D^2} \right)^2 = E_0(t) \left( \frac{\overline{\delta D \, \delta \rho_S}}{\overline{n}} - \delta E_i \right) + \frac{4\pi k_B T}{\overline{n} e} \overline{\delta \rho_S (\delta \rho_S + \delta n)}, \tag{2}$$

$$\Gamma_R \frac{\partial \overline{\delta D \, \delta \rho_S}}{\partial t} + \left( \alpha_R + 3\beta \overline{D}^2 + \beta \, \overline{\delta D^2} \right) \overline{\delta D \, \delta \rho_S} = -E_0(t) \frac{\overline{\delta \rho_S \, \delta n}}{\overline{n}}. \tag{3}$$

$$\Gamma \frac{\partial \overline{\gamma}}{\partial t} + \left( \alpha + 3\beta \, \overline{\delta D^2} + 3\beta \overline{D}^2 \right) \overline{\gamma} = -\left( \alpha_T + 3\beta \, \overline{\delta \gamma} \right) \overline{D}, \tag{4}$$

$$\frac{\Gamma_R}{2} \frac{\partial \overline{\delta \gamma}}{\partial t} + \left( \alpha_R + 2\beta \overline{\delta D^2} + 3\beta \overline{D}^2 \right) \overline{\delta \gamma} = E_0(t) \frac{\overline{\delta \gamma \, \delta \rho_S}}{\overline{n}} - \left( \alpha_{RT} + 6\beta \, \overline{D} \, \overline{\gamma} \right) \overline{\delta D^2} + \frac{4\pi k_B}{\overline{n} e} \overline{\delta \rho_S (\delta \rho_S + \delta n)}, \tag{5}$$

$$\Gamma_R \frac{\partial \overline{\delta \gamma \, \delta \rho_S}}{\partial t} + \left( \alpha_R + \beta \overline{\delta D^2} + 3\beta \overline{D}^2 \right) \overline{\delta \gamma \, \delta \rho_S} = -\left( \alpha_{RT} + \beta \, \overline{\delta \gamma} + 6\beta \overline{D} \, \overline{\gamma} \right) \overline{\delta D \, \delta \rho_S}. \tag{6}$$

The built-in electric field $E_i(l,t) = \frac{4\pi \gamma}{l} \left( \overline{\delta n(t) + \delta \rho_S} \right)_{x,y} \Big|_{-l/2}^{+l/2}$ in (1) is inversely proportional to the film thickness $l$, thus it vanishes in the bulk material [20]. For a finite film it is induced by the space charge layers accommodated near the non-equivalent boundaries $z = \pm l/2$ of examined heterostructure/multilayer (e.g. near the substrate with bottom electrode and free surface with top electrode depicted in the Fig.2). Such layers are created by the screening carries [5], [6] and by possible temporal redistribution of traps or vacancies near the substrate [8]. The built-in field value causes the horizontal shift of hysteresis loops observed in the ferroelectric films with thickness from dozen nanometers up to several microns [26], [27]. In general case the field $E_i(l,t)$ can be time-dependent, its amplitude is proportional to the space charge fluctuations $|\delta n + \delta \rho_S|$. Also $E_i$ diffuses paraelectric-ferroelectric phase transition, in particular it shifts and smears dielectric permittivity temperature maximum. Thus ferroelectric film thickness decrease leads to the degradation of their ferroelectric and dielectric properties [18] up to the appearance of relaxors features [28].

Bratkovsky and Levanyuk [10] predicted the existence of built-in field in a finite ferroelectric film within the framework of phenomenological consideration. The field had a rather general nature. Our statistical approach confirms their assumption and gives the possible expression of the field existing in the inhomogeneous ferroelectric-semiconductor film. The inner electric field induced by the film-substrate misfit strain has been calculated in our recent paper [29]. Misfit-induced field could be the main reason of the loop shift only in the ultrathin strained films with the thickness less than 50 nm.

The renormalization $\Gamma_R \equiv \Gamma + \tau_m$ of Khalatnikov kinetic coefficient in (2), (3), (5), (6) is connected with the contribution of free carrier relaxation with characteristic time $\tau_m = 1/4\pi\mu\overline{n}$ ($\mu$ is $n$-carrier mobility). Usually the time $\tau_m \sim 10^{-6} s$ is much smaller than the initial coefficient $\Gamma > 10^{-3} s$ for the typical concentration of impurity $(0.01 - 1)\%$ and room temperatures [5]. Thus $\Gamma \approx \Gamma_R$.

The renormalization of coefficients $\alpha_R \equiv \alpha + (\gamma + k_B T / 4\pi\overline{n}e)/d^2$ and $\alpha_{RT} \equiv \alpha_T + k_B / 4\pi\overline{n}ed^2$ in (2), (3), (5), (6) is connected with the contribution of correlation and screening effects [21]. Coefficient $\alpha = -\alpha_T (T_C - T)$ is negative in the perfect ferroelectric phase without random defects ($\overline{\delta\rho_S^2} = 0$). For the partially disordered ferroelectric with charged defects ($\overline{\delta\rho_S^2} > 0$) coefficient $\alpha_R$ is positive and $\alpha_R >> |\alpha|$, $\alpha_{RT} >> \alpha_T$. For example, for Pb(Zr,Ti)O$_3$ solid solution $\alpha \sim -(0.4 \div 2)\cdot 10^{-2}$ [3], gradient term $\gamma \approx 5\cdot 10^{-16} cm^2$, screening radius $R_D \sim (10^{-6} \div 10^{-4}) cm$ [5], average distance between defects $d \sim (10^{-6} \div 10^{-4}) cm$ and thus $\alpha_R \sim 1 \div 10^2$. So the ratio $\xi = -\alpha_R/\alpha \approx \alpha_{RT} T / \alpha_T (T_C - T)$ is greater than 100.

The sources of fluctuations in (2), (5) originate from the correlation between diffusion current and displacement fluctuations: $-\overline{\delta D \cdot \partial \delta\rho_S / \partial z} \approx \overline{\delta\rho_S \cdot \partial \delta D / \partial z} \sim \overline{\delta\rho_S (\delta\rho_S + \delta n)}$ [20]. The dimensionless amplitude of these correlations $g = 4\pi k_B T \cdot \overline{n}/(-\alpha D_S^2 e)$ varies in the range from $10^2$ to $10^4$ for Pb(Zr,Ti)O$_3$ ($D_S = \sqrt{-\alpha/\beta}$ is spontaneous displacement of the perfect material with $\overline{\delta\rho_S^2} = 0$). The positive correlator $R^2(t) = -\overline{\delta\rho_S \delta n(t)}/\overline{n}^2$ was calculated in [20] at small external field amplitude and low frequency. In general case correlator $R^2(t)$ is time-dependent, but varies in the range (0; 1) because its amplitude is proportional to the charged defects disordering $\overline{\delta\rho_S^2}\big/\overline{\rho}_S^2$ under the condition of prevailing extrinsic conductivity $\overline{n} \approx -\overline{\rho}_S$ [20].

Hereinafter we discuss only the system pyroelectric response near the equilibrium states. The equilibrium solution of (1)-(6) corresponds to the quasi-static external field changing, e.g. for the harmonic applied field $E_0(t) = E_a \sin(\omega t)$ the ratio $|\Gamma\omega/\alpha|$ should be smaller than unity. At low frequencies $\omega << -\alpha/\Gamma$ the initial conditions do not play significant role in the ferroelectric and pyroelectric hysteresis loops shape. The system (1)-(6) quasi-equilibrium behavior is described by the dimensionless built-in field amplitude $E_m = E_i/E_C$ and frequency $w = -\Gamma\omega/\alpha$ as well as by the aforementioned parameters $\xi$, $R^2(w)$, $g$ and temperature $T/T_C$ ( $E_C = -\alpha\sqrt{-\alpha/\beta}$ is proportional to the thermodynamic coercive field of the perfect material with $\overline{\delta\rho_S^2} = 0$ ). Keeping in mind that carrier fluctuations $\delta n(t)$ are time-dependent, their frequency spectrum $\delta n(w)$ determines the frequency dispersion of parameters $R^2(w) \sim \overline{\delta n \delta\rho_S}$ and $E_m(w) \sim (\delta n + \delta\rho_S)$. It is appeared, that under the conditions $w < 1$, $g >> 1$ and $\xi >> 1$ the scaling parameter $gR^2/\xi$ determines the system behavior.

Figs 3a demonstrate the typical changes of pyroelectric hysteresis loop caused by the increase of charged defects disordering (note, that $gR^2/\xi \sim \overline{\delta\rho_S^2}\big/\overline{\rho}_S^2$ ).

Figs 3b,d demonstrates, that disorder parameter $\overline{\delta\gamma} \sim \overline{\delta D^2}$ exists in the all region of pyroelectric hysteresis and reaches its maximum value near the coercive field, where $\overline{D} \to 0$ and $\overline{\gamma} \to 0$. Let us underline, that in contrast to Landau-Khalatnikov equations describing homogeneous reversal with $\overline{\delta D^2} \equiv 0$ near the coercive field, our coupled equations demonstrate **inhomogeneous** ferroelectric and pyroelectric response reversal [20]. Namely when the external field reaches the coercive value the sample splits into the oppositely polarized regions, so that it is non-polarized as a whole. Every polarized region causes pyroelectric coefficient fluctuations $\delta\gamma(\mathbf{r}, t)$.

It is clear from the Figs 3a,c, that the increase of $gR^2/\xi$ value leads to the essential decrease and smearing of pyroelectric coefficient peaks near coercive field and to the decrease of the coercive field value (compare Landau-Khalatnikov loops ( $R^2 = 0$ ) with dashed curves ( $R^2 \geq 0.2$ )). At $R^2 \geq 0.5$ pyroelectric coefficient peaks near coercive field completely disappears and typical pyroelectric hysteresis loop looks like the ferroelectric one. At $R^2 \geq 0.8$ the coercive field is much smaller than its thermodynamic value at $R^2 = 0$. Thus, pyroelectric loop becomes sloped, much thinner and little lower under the increase charged defects disordering $\overline{\delta\rho_S^2}$ . This effect is similar to the well-known **"square to slim transition"** of the ferroelectric hysteresis loops in relaxor ferroelectrics [14].

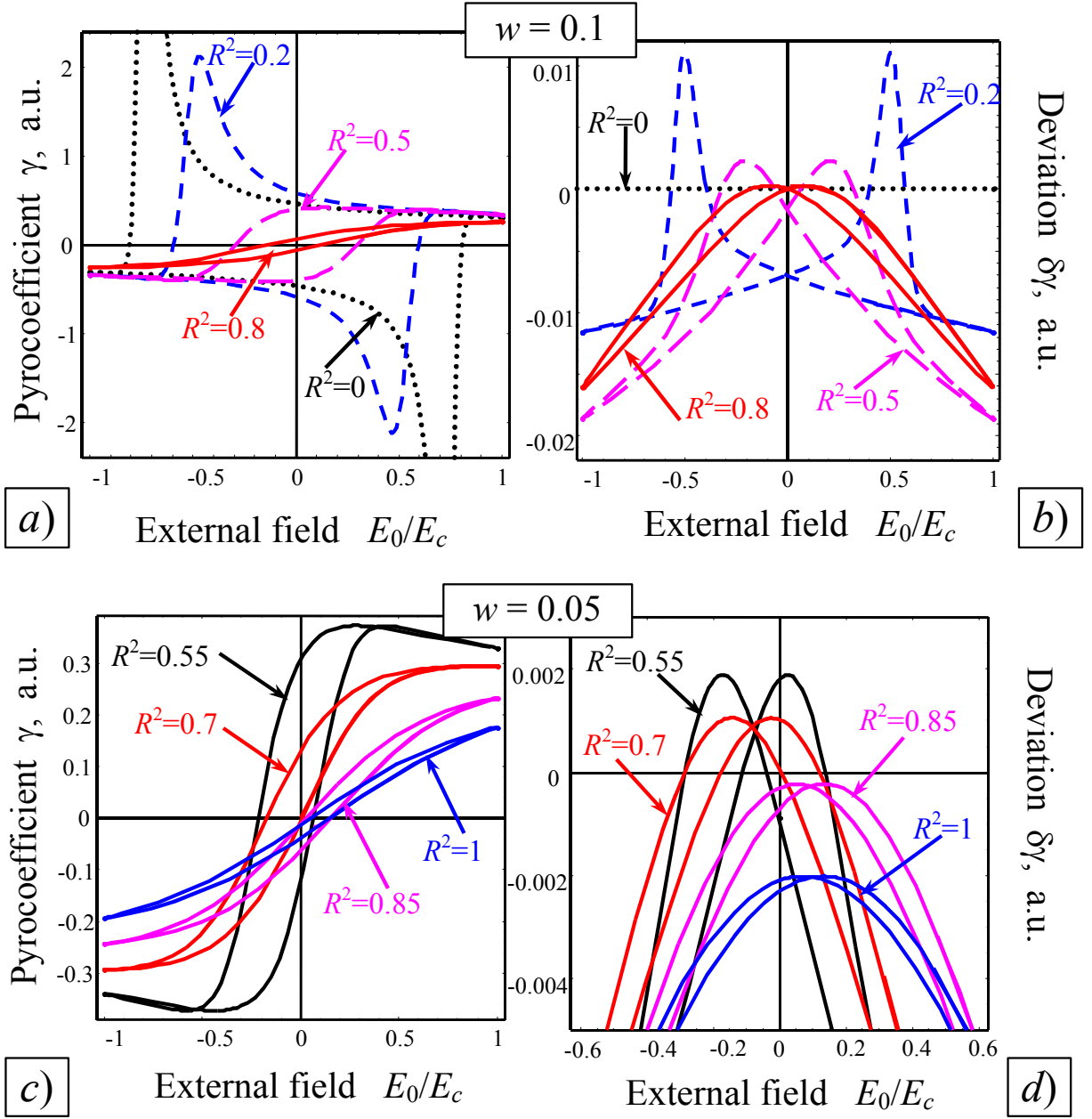

**Figure 3.** Hysteresis loops of pyroelectric coefficient $\overline{\gamma(E)}$ and its deviation $\overline{\delta\gamma(E)}$ for different $R^2$ values. Other parameters: $g$=100, $\xi$=100, $T/T_C$=0.45, $E_i = 0$, $w$=0.1 (plots (a, b) for the bulk sample) and $E_m = \pm 0.1 \cdot R$, $w$=0.05 (plots (c,b) for the film).

It seems that the hysteresis loop width could only decrease, when the donor concentration increases, because the width decreases under $gR^2/\xi$ increasing. However the parameter $gR^2/\xi$ decreases when acceptors (e.g. intrinsic vacancies or traps) compensate donors. The loop width increases under $gR^2/\xi$ decreasing. Such compensation is possible in "hard" $Pb(Zr,Ti)O_3$ (PZT) ceramics, where donor impurities do not change the loops width [3].

Let us underline, that we do not know any experiment, in which pyroelectric coefficient peaks near coercive field have been observed. Moreover, usually pyroelectric hysteresis loops in doped ferroelectrics have typical "slim" shape with coercive field values much lower than the thermodynamic one [30], [31]. So, our approach qualitatively explains available experimental results. The quantitative comparison of our numerical modelling with typical PZT-pyroelectric loops is presented in the next section.

### 3. COMPARISON WITH EXPERIMENTAL RESULTS AND DISCUSSION

Dopants, as well as numerous unavoidable oxygen $O^{-2}$ vacancies, can play a role of randomly distributed charged defects in "soft" PZT. Really, it is known that low level (0.001-1.0 %) donor additives La, Nb, Nd or Ce soften PZT dielectric and pyroelectric properties at room temperature. Ferroelectric and pyroelectric hysteresis loops have got relatively high $\gamma$ and $D$ remnant values, but reveal low coercive fields in soft PZT [3]. Usually pyroelectric hysteresis loops of PZT are rather slim and sloped even at low frequencies $\omega \sim (0.1 \div 10)Hz$ [3], any pyroelectric coefficient maximum near the coercive field is absent [30], [31].

Investigated Pt-PZT-Pt/Ti-SiO$_2$/Si structures with oriented PZT layer were manufactured by radio frequency magnetron sputtering in the system under the conditions described previously [17]. The sputtering target obtained by uniaxial cold pressing includes the mixture of PbO, TiO$_2$ and ZrO$_2$ in a stochiometric composition. The structure includes the top 150 nm Pt- electrode, 1.9 μm layer of oriented PZT, bottom Pt/Ti-bilayer (150 nm of Pt, 10 nm of Ti) deposited onto the oxidized (350 nm of SiO$_2$) (100) n-type Si 350 μm substrate.

For PZT − Si-substrate structure it is necessary to design the bottom electrode, which possesses not only a stable and high enough electrical conductivity but also simultaneously prevents the interfacial reactions between electrode, PZT and SiO$_2$ components in PZT-film and Si-substrate surroundings under rather high temperature. The layer of Ti plays an important role in limiting the diffusion of Ti in Pt/Ti intermediate bilayer through Pt-layer into the PZT-layer and directly into SiO$_2$-layer, and also in correction of poor adhesion of Pt-layer. The annealing treatment of the Pt/TiO$_x$/SiO$_2$/Si-substrate structure just before of PZT deposition was performed for substrate stabilization and post-annealing treatment of PZT-film was performed for crystallizing the film in the polar perovskite phase. The top Pt- electrode has 1 mm$^2$ area.

The pyroelectric response of the PZT films was registered by means of dynamic pyroelectric measurements (see [30] for details). During the measurements the quasi-static voltage $V$ varied in the range (-11V, +11V) at the low-frequency $\omega \sim 0.01 \, Hz$, the temperature $T$ changes near the room one with the frequency about 20 Hz. Pyroelectric hysteresis loops for $U_{\pi 1}(V) \sim \overline{\gamma}$ and $U_{\pi 2}(V) \sim \overline{\gamma}/\overline{\varepsilon}$ were reconstructed from pyroelectric response amplitude $\left| U_{\pi 1,2}(V) \right|$ and phase $\varphi_{\pi 1,2}(V)$. Experimental loops and our calculations are presented in the Figs.4.

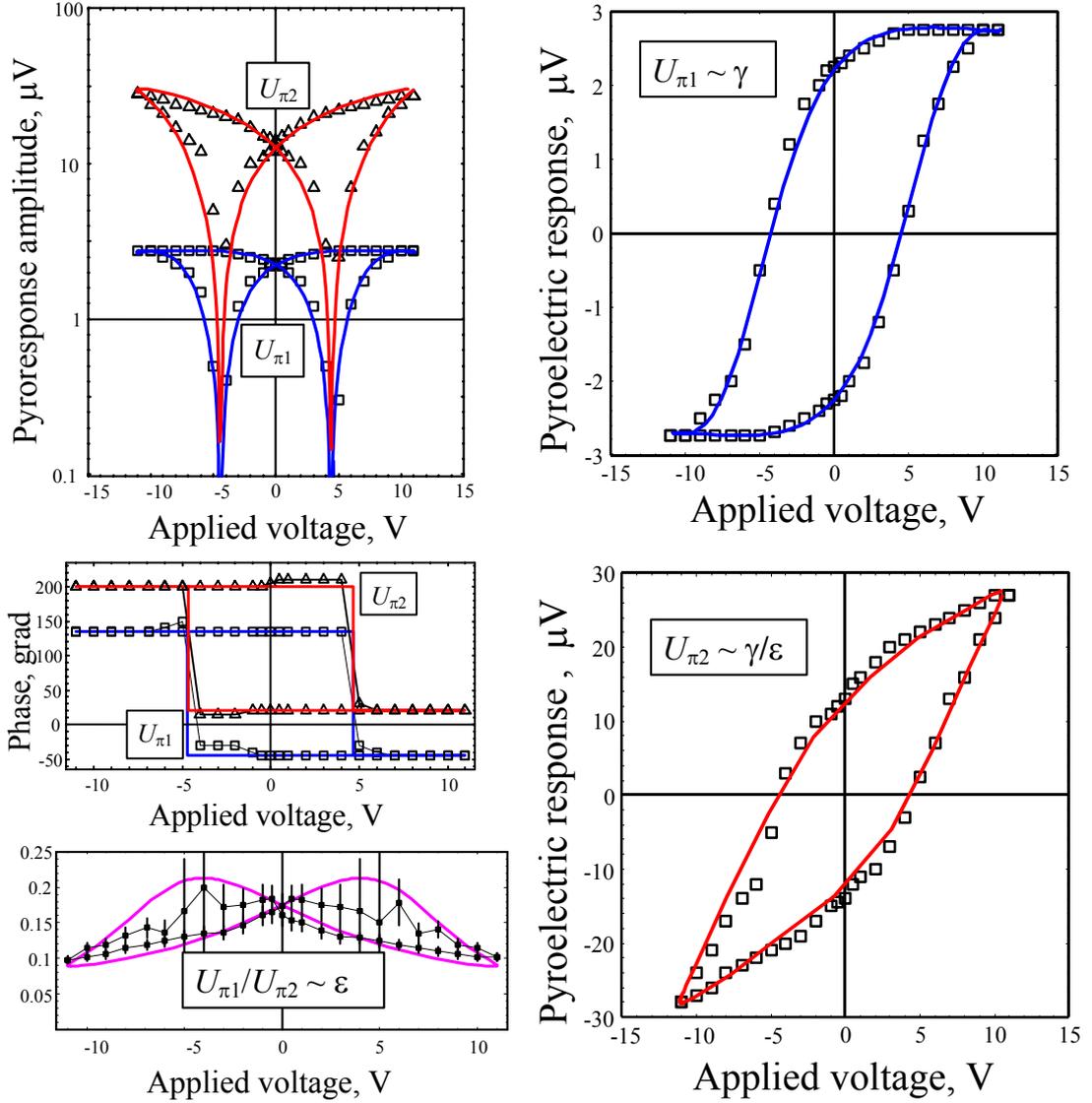

**Figure 4**. Pyroelectric response of 1.9 μm-thick PZT(46/54):Nb film. Pyroelectric hysteresis loops (right plots for $U_{\pi 1} \sim \overline{\gamma}$ and $U_{\pi 2} \sim \overline{\gamma}/\overline{\varepsilon}$) have been reconstructed from pyroelectric response amplitude and phase (left plots for $\left| U_{\pi 1,2} \right|$ and $\varphi_{\pi 1,2}$). Squares are experimental data measured by Bravina *et al.* [30], solid curves are our calculation with the fitting parameters $w$=0.1, $R^2$=0.5, $g$=100, $\xi$=100, $E_m$= -0.03.

It is clear from the figures, that our model both qualitatively and quantitatively describes pyroelectric hysteresis loops in thick "soft" PZT films. Our modelling of ferroelectric and dielectric hysteresis loops was performed earlier (see e.g. [20]). Ferroelectric and dielectric hysteresis loops measured for the same PZT films by using conventional Sawyer-Tower method and impedance analyzer are presented in Figs 5. Note, that a film capacity is proportional to the dielectric permittivity $\overline{\varepsilon}$ (see equations for $\overline{\varepsilon}$ in Appendix B).

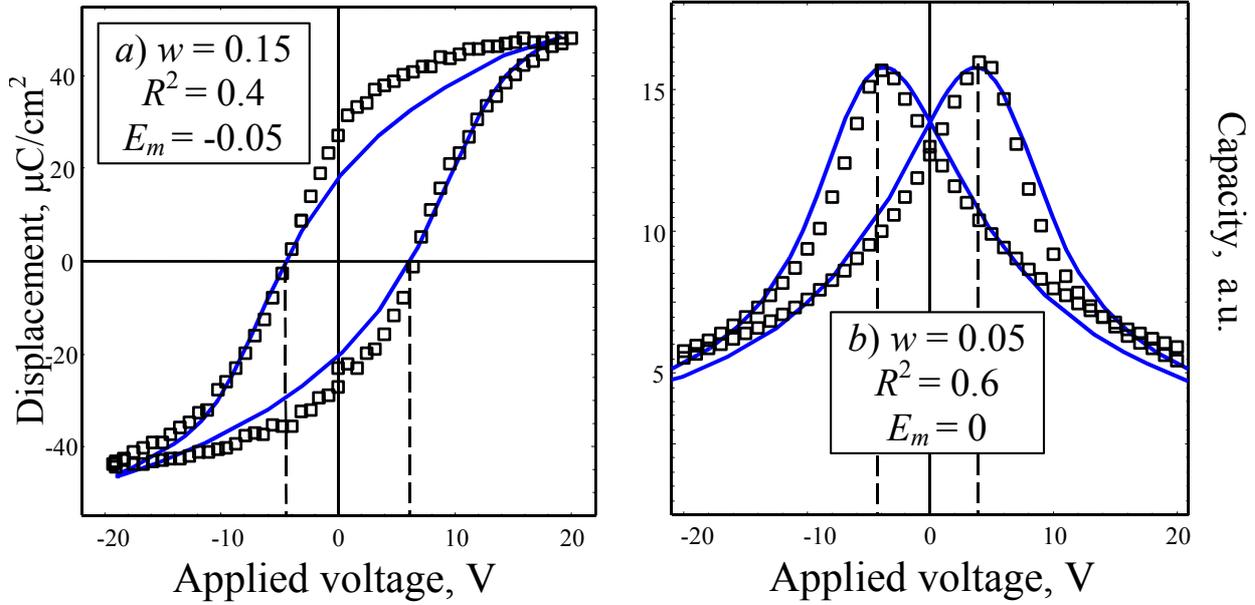

**Figure 5.** Ferroelectric (a) and dielectric (b) hysteresis loops of 1.9 μm-thick PZT(46/54):Nb film. Squares are experimental data measured at room temperature, solid curves are calculations with the fitting parameters $g$=100, $\xi$=100 and different built-in fields $E_m$, frequencies $w$ and disordering $R^2$. All other parameters correspond to the ones for perfect PZT solid solution [3].

It is clear from the Figs 4-5, that ferroelectric loop is slightly asymmetrical, i.e. built-in electric field $E_i \neq 0$, whereas the pyroelectric and dielectric loops measured at lower frequency look almost symmetrical. This means that built-in field is dynamic in the aforementioned experiments: its value is noticeable at higher external field frequency $w = 0.15$ and negligibly small at low frequency $w = 0.05$. Note, that sometimes even stabilized loop (i.e. obtained after many circles) is asymmetrical [29], [31]. Also we obtained that disordering $R^2(w) \sim \overline{\delta\rho_S \delta n(w)}$ decreases with external field frequency increase, namely it varies from 0.6 to 0.4 under the frequency change from 0.05 to 0.15 (see Figs 4, 5). This result could be explained allowing for the fact, that the defect-carrier dipole coupling influences the switching mechanism [8], but dipole correlations "sluggish defect $\delta\rho_S$ + mobile fluctuation $\delta n$" weaken with the frequency increase.

Thus, the modelling based on the coupled equations (1)-(6) gives realistic coercive field values and typical pyroelectric hysteresis loop shape, in contrast to the Landau-Khalatnikov approach, that describes the homogeneous pyroelectric response reversal. Taking into account that the inhomogeneous reversal of spontaneous polarization and pyroelectric response occurs in the doped or inhomogeneous ferroelectrics-semiconductors, the proposed coupled equations can be more relevant for the phenomenological description of their polar and pyroelectric properties, than the models based on Landau-Khalatnikov phenomenology.

## APPENDIX A

The expressions for correlations $\overline{\delta D \delta E}_z$, $\overline{\delta \rho_s \delta E}_z$ and $\gamma \overline{\dfrac{\partial^2 \delta D}{\partial \mathbf{r}^2}}$ were derived in [20] for the bulk sample.

For a finite film with thickness $l$ they acquire the form:

$$\gamma \overline{\frac{\partial^2 \delta D}{\partial \mathbf{r}^2}} = \gamma \overline{\left(\frac{\partial \delta D}{\partial z}\right)}_{x,y} = \frac{4\pi\gamma}{l} \overline{(\delta n + \delta \rho_S)}_{x,y} \Big|_{-l/2}^{+l/2} \tag{B.1}$$

$$\overline{\delta D \delta E}_z = -\frac{1}{8\pi\mu\overline{n}} \frac{\partial}{\partial t} \overline{\delta D^2} + E_0(t) \frac{\overline{(\delta\rho_S \delta D)}}{\overline{n}} - \frac{E_0(t)}{8\pi l\,\overline{n}} \left(\overline{\delta D^2}\right)_{x,y}\Big|_{-l/2}^{+l/2} - \frac{\kappa}{\mu\overline{n}l} \left(\overline{\delta D \delta\rho_S}\right)_{x,y}\Big|_{-l/2}^{+l/2} +$$

$$+ \frac{\kappa}{\mu}\overline{\left(\frac{\delta\rho_S}{\overline{n}}(\delta\rho_S + \delta n)\right)} - \frac{R_D^2}{d^2}\overline{\delta D^2} \tag{B.2}$$

$$\overline{\delta\rho_s \delta E}_z = -\frac{1}{4\pi\mu\overline{n}} \frac{\partial}{\partial t}\overline{\delta\rho_S \delta D} - E_0(t)\frac{\overline{\delta\rho_S \delta n}}{\overline{n}} - \frac{\kappa}{2\overline{n}\mu l}\left(\overline{\delta\rho_S^2}\right)_{x,y}\Big|_{-l/2}^{+l/2} + \frac{\kappa}{4\pi\,\overline{n}\,\mu\,d^2}\overline{\delta\rho_S \delta D} \tag{B.3}$$

If the charged defects distribution is quasi-homogeneous, expressions (B.1-3) can be simplified, namely

$$\frac{4\pi\gamma}{l}\overline{(\delta n + \delta\rho_S)}_{x,y}\Big|_{-l/2}^{+l/2} \approx \frac{4\pi\gamma}{l}\left(\overline{\delta n}\right)_{x,y}\Big|_{-l/2}^{+l/2}, \quad \frac{\kappa}{\mu\overline{n}l}\left(\overline{\delta D \delta\rho_S}\right)_{x,y}\Big|_{-l/2}^{+l/2} \approx 0, \quad \frac{\kappa}{2\overline{n}\mu l}\left(\overline{\delta\rho_S^2}\right)_{x,y}\Big|_{-l/2}^{+l/2} \approx 0, \tag{B.4}$$

$$\frac{\left(\overline{\delta D^2}\right)_{x,y}}{8\pi l\,\overline{n}}\Big|_{-l/2}^{+l/2} = \frac{2\pi}{l\,\overline{n}} \overline{\left(\int_{z_0}^{z} dz(\delta n + \delta\rho_S)\right)^2}_{x,y}\Big|_{-l/2}^{+l/2} \sim \overline{(\delta n + \delta\rho_S)^2} \tag{B.5}$$

The constant $z_0$ can be determined from the boundary conditions for $\overline{\delta D}_{x,y}$ at $z$=0. Under the condition $z_0 = \pm l/2$, built-in field deviation $\delta E_i$ is absent.

## APPENDIX B

The system determining the dielectric permittivity $\overline{\varepsilon} = \partial\overline{D}/\partial E_0$, its deviation $\overline{\delta\varepsilon} = \partial\overline{\delta D^2}/\partial E_0$ and correlation $\overline{\delta\varepsilon\delta\rho_S} = \partial\overline{\delta D \delta\rho_S}/\partial E_0$ has the form:

$$\Gamma\frac{\partial\overline{\varepsilon}}{\partial t} + \left(\alpha + 3\beta\overline{\delta D^2} + 3\beta\overline{D}^2\right)\overline{\varepsilon} = 1 - 3\beta\ \overline{\delta\varepsilon\overline{D}}, \tag{A1}$$

$$\frac{\Gamma_R}{2}\frac{\partial\overline{\delta\varepsilon}}{\partial t} + \left(\alpha_R + 2\beta\overline{\delta D^2} + 3\beta\overline{D}^2\right)\overline{\delta\varepsilon} = -\frac{\overline{\delta D\delta\rho_S}}{\overline{\rho}_S} - E_0(t)\frac{\overline{\delta\varepsilon\delta\rho_S}}{\overline{\rho}_S} - 6\beta\ \overline{\delta D^2}\ \overline{D}\ \overline{\varepsilon}, \tag{A2}$$

$$\Gamma_R\frac{\partial\overline{\delta\varepsilon\delta\rho_S}}{\partial t} + \left(\alpha_R + \beta\overline{\delta D^2} + 3\beta\overline{D}^2\right)\overline{\delta\varepsilon\delta\rho_S} = \frac{\overline{\delta\rho_S\delta n}}{\overline{\rho}_S} - \beta\left(\overline{\delta\varepsilon} + 6\overline{D}\ \overline{\varepsilon}\right)\ \overline{\delta D\delta\rho_S}. \tag{A3}$$